\documentstyle[12pt]{article}
\topmargin- 23mm
\begin{document}
\vspace*{.5cm}
\begin{center}
{ \bf ON ELECTROWEAK MOMENTS OF BARYONS AND SPIN--FLAVOUR STRUCTURE OF THE
NUCLEON}
\footnote{This work was supported by the RFBR grants No.96-02-18137 and
No.96-15-96423}\\
\vspace*{.3cm}

S.B.Gerasimov \\
\vspace{.1cm}

{\it Bogoliubov Laboratory of Theoretical Physics, JINR, Dubna}
\end{center}

\vspace*{.5cm}

\baselineskip 20pt
The phenomenological sum--rule--based approach is used to discuss the quark
composition dependence of some static and quasi--static electroweak
characteristics of nucleons.The role of nonvalence degrees of freedom,
the nucleon sea partons and/or peripheral meson currents, is shown
to be important to select and make use of the relevant symmetry
parametrization of hadron observables. With our preferable universal value
of the $SU(3)$-symmetry parameter $\alpha_{D}=D/F+D=.58$,
taken for both magnetic moments and axial-vector constants entering
into the semi-leptonic baryon decays, we obtain the following values for
moments $\Delta q$ of the spin-dependent structure function of the proton:
$\Delta u \simeq .84(.82), \Delta d \simeq  -.42(-.44), \Delta s=-.22 \pm .05
(-.10 \pm .03)$, where the values in parentheses correspond to the widely
used "standard" value of $\alpha^{axial}_{D}=.63$. The estimations of the
strange sea contributions to the nucleon magnetic moments and rms  are also
presented.\\
\vspace*{.5cm}

{\bf 1.} The magnetic moments of the lowest baryon octet, being one of the most
accurately measured spin--dependent quantities in hadron physics \cite{a_1},
may serve as a useful means to verify the spin--flavour symmetry predictions
and different model calculations \cite{a_2,a_3,a_4,a_5} of the nucleon
structure characteristics.
This report aims at discussing, on the basis of confrontation of data on the
magnetic moments and axial--vector coupling of baryons, some alternative,
to the usually discussed, inferece for the $\Delta q$'s of the polarized DIS
and the (hidden) strangeness--dependent characteristics of the nucleon.
In  \cite{a_3,a_4} (and the references to earlier works therein),
the following parametrization was introduced for magnetic moments
$\mu(B)$ of baryons :
\begin{eqnarray}
&& \mu(B)=\mu(q_e) g_2 + \mu(q_o) g_1 + C(B) + \Delta, \label{e1}\\
&& \mu(\Lambda)=\mu(s) ({2\over 3} g_2 -{1\over 3} g_1) + (\mu(u) + \mu(d))
   ({1\over 6} g_2 + {2\over 3} g_1) + \Delta, \label{e2}\\
&& \mu(\Lambda\Sigma)={1\over \sqrt{3}} (\mu(u) - \mu(d))
   ({1\over 2} g_2 - g_1) + C(\Lambda\Sigma), \label{e3}\\
&& \Delta = \sum\limits_{q=u,d,s} \mu(q) \delta(N),\label{e4}
\end{eqnarray}
where $\mu(q)$ are the effective quark magnetic moments defined without
any nonrelativistic approximations,
$g_{2(1)}$ are "reduced" dimensionless
coupling constants obeying exact $SU(3)$--symmetry and related with the SU(3)
$F-$ and $D-$ type constants via $g_2=2F$ and $g_1=F-D$, $\Delta(B)$ is a
matrix element of the OZI--suppressed $\overline qq$--configuration for
a given hadron: $\Delta(B)=<B|\overline qq |B>$, where $q\not\subset
\{q_e^2, q_o\}$, e.g. $\delta(N)=<N|\overline ss |N>$, etc.
The $C(P)=-C(N)$ and $C(\Lambda\Sigma)$ are representing
the isovector contributions
of the charged--pion exchange current to $\mu(P), \mu(N)$ and the
$\Lambda\Sigma$--transition moment $\mu(\Lambda\Sigma)$.
In Refs. \cite{a_3,a_4} the use was made of two pictures of the baryon
internal composition.In the first one, all baryons are considered
as consisting of three massive, "dressed" constituent quarks, locally
coupled with lightest goldstonions -- the pseudoscalar octet fields.
In the second picture
only fundamental QCD quanta, the quarks and gluons, are there, the
meson component of the baryon state vectors being represented by the
properly correlated "current" quarks and gluons.The use of one picture or
another will be reflected in a particular parametrization of
contributions due to corresponding nonvalence degrees of freedom.
Here, we concentrate on two of the earlier discussed \cite{a_3,a_4} sum rules
( we use the particle and quark symbols for corresponding
magnetic moments):
\begin{eqnarray}
 \alpha ={D\over F+D} = {g_2 - 2g_1\over 2(g_2-g_1)} = {1\over 2}
   \left(1 - {\Xi^0 - \Xi^-\over \Sigma^+ - \Sigma^- -
   \Xi^0 + \Xi^-} \right),\label{e5}\\
 {u\over d} = \frac{\Sigma^+ (\Sigma^+ - \Sigma^-) - \Xi^0 (\Xi^0 - \Xi^-)}{\Sigma^- (\Sigma^+ - \Sigma^-) - \Xi^- (\Xi^0 - \Xi^-)} = \label{e6} \\
            = \frac{P + N + \Sigma^+ - \Sigma^- + \Xi^0 - \Xi^-}{P + N - \Sigma^+ + \Sigma^- - \Xi^0 + \Xi^- },\label{e7}
\end{eqnarray}
Eqs.(6,7) were obtained provided $\delta(N)=0$, hence they are related to
the chiral constituent quark model where a given baryon consists of three
"dressed", massive constituent quarks.
Eqs.(6,7) also show that owing to the virtual transitions
$q \leftrightarrow \pi (\eta) + q$, $q \leftrightarrow K + s$
the "magnetic anomaly" is developing, i.e. $u/d=-1.80\pm 0.02 \not=
Q_u/Q_d = -2.$
Evaluation of the one--loop, quark--meson diagrams
gives : $u/d=(Q_u +\kappa_u)/(Q_d+\kappa_d) \simeq -1.85$, the $\kappa_q$
being the quark anomalous magnetic moment in natural units,
if we take the $SU(3)$ - invariant quark-pseudoscalar- meson couplings,
the physical masses for the $\pi-, \eta-, K $--mesons  and
the $m_{q(s)} \simeq 300 (460)$ MeV.
In the $SU(3)$--limit, when $m_q=m_s$ and $m_{\pi}=m_{\eta}=m_{K}$,
we return to $u/d = - 2$, the ratio pertinent to the structureless
"current" quarks.\\
Normalizing the "strange charge" coupling constant for the $s$--quark
($K$--meson)
to - $\frac{1}{3} (+\frac{1}{3})$, i.e. the values coinciding with
the electric charge of the $s$-- and $\overline s$--quarks, we obtain,
in the one--loop appoximation, the contribution of strange quarks to
the anomalous magnetic moments (a.m.m.) of the $u$-- and $d$-- constituent
quarks, which also coicides approximately with the corresponding
contribution to nucleon magnetic moments
\begin{eqnarray}
&& \kappa_{u(d)}(s,\overline s) \simeq \mu_{P(N)}(s,\overline s) = 0.065 n.m. \label{8}
\end{eqnarray}
Recalling the negative electric charge of the strange quark, we conclude
that the spin of strange quark appearing inside the polarized nonstrange
quark is antiparallel to that of the "parent" constituent $u$-- or $d$--quark.
The average polarization of the $\overline s$--quark, forming the $K$--meson,
is zero.\\
With the normalization of the metioned coupling constants to the values
$+1 (-1)$, we obtain the (hidden) strange quark contributions to
the Dirac "strange charge" radius and the a.m.m. of nonstrange quarks
the values:
\begin{eqnarray}
& \kappa_{u(d)}(s,\overline s) \simeq - 0.195 n.m. \nonumber \\
& <r_{1}^2>_{u(d)}(s,\overline s) \simeq \frac {1}{3}<r_{1}^2>_{P(N)}(s,\overline s) \simeq 0.013 fm^2 \label{9}
\end{eqnarray}
We stress, however, that the virtual $K$--mesons were treated as
point--like particles.The intrinsic strange antiquark distribution
in the virtual $K$-- mesons may not be negligible, e.g. for the on--shell
$K$--mesons, with the adopted above--mentioned normalization, one can obtain
the estimate:
$$ <r^2>_{K}(\overline s) = - <r_{ch}^2>_{K^{+}} - 2<r_{ch}^2>_{K^{o}} = - 0.3 fm^2$$
where the one--loop calculation for $<r_{ch}^2>_{K^{+}(K^{o})}$ were taken
from \cite{a_6}.
Therefore even the sign of the whole value of $<r_{1}^2>_{P(N)}(s,\overline s)$
may be reversed after a proper inclusion of the $<r^2>_{K}(\overline s)$
as a part of a still missing two--loop calculation.\\
With the neglect of the nonvalence contribution, i.e. with $C(B's) =
\Delta(B's) = 0, u/d=-2$ we obtain for magnetic moments of baryons
the results coinciding almost identically with the
results of the $SU(6)$--based nonrelativistic quark model, taking account
of the $SU(3)$-- breaking due to the quark--mass differences \cite{a_7}.
We stress, however, that no NR assumption or explicit $SU(6)$-wave function
are used this time.\\
The ratio $\alpha_{D}$ equals 0.61 in this case( cf. Eq.(5), giving
the value 0.57 ), thus demonstrating a substantial influence of
the nonvalence ( i.e. the meson ) degrees of freedom on this important
parameter.\\
We turn now to a complementary view of the nucleon structure, absorbing
$C(N's)$ into products of the corresponding $\mu(q)$ and $g(N's)$,
keeping the constraint $u/d = -2$,and $\Delta(B's)$ non-zero.
We shall refer to this approach \cite{a_3,a_4} as
a correlated current quark picture of nucleons. In this
case we have
\begin{eqnarray}
&& \Delta(N) = \sum\limits_{q=u,d,s} \mu(q) \delta(N)
= {1\over 6}(3(P+N) - \Sigma^+  + \Sigma^- -\Xi^0 +\Xi^-) = \nonumber \\
&& -.062\pm.01 n.m.,\label{9}\\
&& \mu_N(\overline ss)=\mu(s)
   \langle N|\overline ss|N\rangle=
   (1- {d\over s})^{-1}\Delta =.11 n.m. ,\label{10} \\
&& (\alpha_D)_N = \frac {3(N - \Delta(N))}{2(N - P)} = 0.59,\label{11}
\end{eqnarray}
where independence of the sum $P+N$ of the $C(N's)$ and the ratio
$d/s=1.55$ \cite{a_3,a_4} have been used.\\
By definition, $\mu_N(\overline s,s)$ represents the contribution of
strange ("current") quarks to nucleon magnetic moments.Numerically, Eq.(11)
agrees fairly well with other more specific models ( see,e.g. \cite{a_8})
but exceeds the value given by Eq.(8) by factor of about 2, indicating on
other possible mechanisms of the transition $q \leftrightarrow q + s +
\overline s$. The different values $(\alpha_D)_{N} = 0.59$ and
$(\alpha_D)_{Y} = 0.57$, Eq.(5), give the hint of the difference of
the nucleon and hyperon wave functions that would lead to their
mismatch and corresponding influence on the $Y \to Nl\nu$ -transitions.
Even bigger difference of $\alpha_{D}^{mag} = 0.59$ or $0.57$ and
the average value $\alpha_{D}^{axial} = .635$ also lead to difficulty,
when interpreted
\cite{a_9}
via the admixture of the higher representation
of the $SU(6) \otimes O(3)$ -- basis functions to the ground state.
In that case we would have an unacceptably large admixture,
with the probability of about 0.2, of the D--wave in the nucleon wave
function ( see the next section).\\
In this respect we wish here to recall a less popular inference from
the semi--leptonic hyperon decays and its impact on a description
of the polarized DIS.\\
As is known \cite{a_10}, to obtain the contributions of the $u-,d-$
and $s$-flavoured quarks to the proton
spin, denoted by $\Delta u(p), \Delta d(p)$ and $\Delta s(p)$, the use is
usually made of baryon semileptonic weak decays
treated with the help of the exact $SU(3)$-symmetry. It has been shown
earlier \cite {a_11,a_12} that when both the strangeness-changing
$(\Delta S=1)$ and strangeness-conserving $(\Delta S=0)$
transitions are taken for the analysis,
then $(D/F+D)_{ax}^{\Delta S=0,1} = .635 \pm .005 $ while
$(D/F+D)_{ax}^{\Delta S=0}=.584 \pm .035 $, which is close to the mean value
$ (D/D+F)_{mag} \simeq .58 $, according to Eqs.(5,11).
We list below two sets of the $\Delta q$-values,
we have obtained from the data
with inclusion of the QCD radiative corrections
(e.g.\cite{a_10} and references
therein) : $\Delta u(p) \simeq .82 (.84),\;
\Delta d(p) \simeq -.44(-.42),\;
\Delta s =-.10 \pm .03(-.22 \pm .05)$, where the values in
the parentheses correspond to $\alpha _{D}=(D /D+F)=.58$.
At the same time, the problem of
difference of the following two expressions
\begin{eqnarray}
&& F - D = (g_a/g_v)^{exp}(\Sigma^{-}\rightarrow N) = -.34 \pm .02,\label{e12)}\\
&& F - D = (g_a/g_v)^{exp}(N \rightarrow P) - \sqrt6
g_{a}^{exp}(\Sigma \rightarrow \Lambda)= -.19 \pm .04,\label{e13}
\end{eqnarray}
of which we prefer the second one
when we postulate $\alpha_{D} ^{ax} =\alpha _{D}^{mag}$,
remains open.The intriguing possibility can, however, be mentioned
that the numerical value of the $(g_a/g_v)^{exp}(\Sigma^{-}\rightarrow N)$,
coinciding with Eq.(14), was in fact found in \cite{a_13},
if the "weak--electricity" form factor, $g_{w.el} \neq 0$, referred as one of
the second class current effects, is included in the joint analysis of all
experimental data.\\

{\bf 2.} In this section, we consider the possible difference between
$\alpha_m$, Eq. (\ref{e5}), and the
$\alpha_{ax}$'s, given in \cite{a_11,a_12}, as originating from the
higher $SU(6)\times O(3)$ three--quark configurations and/or the exotic
$(3q+g)$--admixture in the ground state wave function:
\begin{eqnarray}
\Psi &=& A_0\Psi_0 (\{56\}_S, L_q=0, S_q=1/2)
         + A_1\Psi_1 (\{70\}_M, L_q=0, S_q=1/2) \nonumber\\
     &+& A_2\Psi_2 (\{70\}_M, L_q=2, S_q=3/2)
         + A_3\Psi_3 (\{20\}_A, L_q=1, S_q=1/2) \nonumber\\
     &+& A_g\Psi_g (\{3q\}_{8c} + g_{8c}).
\label{e14}
\end{eqnarray}
The coefficients $A_i$ and $A_g$ satisfy the normalization condition
\begin{eqnarray}
\sum\limits_{i=0}^3 A_i^2 + A_g^2 =1.
\label{e15}
\end{eqnarray}
In Eq. (\ref{e14}),  $L_q (S_q)$ is the quark orbital (spin) moment, and
the index "8c" stands for the color--octet states. To specify different cases,
we keep for the gluon angular momentum two simplest possibilities
$J_g^P =1^\pm$ which are the $M1$-- or $E1$-- gluon modes. Different
components of the total wave function are built of the antisymmetrized
products of the flavor ($\Phi$), spin ($\chi$) color ($\omega$) and
orbital/radial ($\Psi$) wave functions:
\begin{eqnarray}
\Psi = \Phi \times \chi \times \omega \times \Psi(\vec\rho, \vec\lambda),
\label{e16}\end{eqnarray}
$\vec\rho$ and $\vec\lambda$ being the Jacobi coordinates of the 3--quark
system. Of many considered possibilities for $\Psi_g$ we present two
examples, one for the $M1$-- and one for the $E1$-- gluon mode
(the ${M1}$-case has been considered also in\cite{15}, but with all higher
orbital configurations neglected, $A_{i}=0$, $i=1,2,3$ )
\begin{eqnarray}
&& \Psi_g^{M1} = {1\over 2} [(\Phi^\rho \omega^\rho -
   \Phi^\lambda \omega^\lambda)\chi^\lambda +
   (\Phi^\rho \omega^\rho + \Phi^\lambda \omega^\lambda)\chi^\rho]
   \Psi_{sym},\label{e17} \\
&& \Psi_g^{E1} = {1\over 2} [(\Phi^\rho\chi^\rho+\Phi^\lambda\chi^\lambda)
    (\omega^\rho\Psi^\lambda - \omega^\lambda \Psi^\rho)], \label{e18}
\end{eqnarray}
where $\Psi_{sym}(L_\rho=L_\lambda=0)=\Psi_0(\rho^2 + \lambda^2), \
\Psi^{\rho(\lambda)} (L_{\rho(\lambda)}=1, L_{\lambda(\rho)}=0) =
\vec\rho(\vec\lambda)\cdot \Psi_1(\rho^2 +\lambda^2), \ \Psi_{0,1}
(\rho^2 + \lambda^2)$ are unspecified radial wave functions,
$\Phi^{\rho(\lambda)}, \omega^{\rho(\lambda)}$, etc  are familiar,
octet--type wave functions (see, e.g., Ref. \cite{14} and earlier
citations therein). Then we find expectation values of the magnetic
moment ($\hat\mu$) and axial charge ($\hat A$) operators
\begin{eqnarray}
&& \hat \mu=\sum\limits_q [g_\sigma(q) \hat\sigma_3(q) +
    g_l(q) \hat l_3(q)],\label{e19}\\
&& \hat A =\sum\limits_q g_{ax}(q) \hat\sigma_3(q)\hat\tau_3(q), \label{e20}
\end{eqnarray}
and define $g_i^m$ in Eq. (\ref{e1})  and the analogous $g_i^{ax}$ as
function of $A_0,...,A_g, g_\sigma, g_l$ and $g_{ax}$. Then we take
the ratios $\alpha_m$ and $\alpha_{ax}$; from the latter the unknown (due to
various renormalization effects) $g_{ax}$ will be cancelled out.
First, we indicate what physics' factors will cause
deviation of $F/D_{ax}$-ratio from $2/3$ ( the $SU(6)$-value ) in either
of two options
\begin{equation}
\frac{F}{D}|_{ax}=\Biggl \{ \begin{array}{llll}
0.58 & \mbox{ if } &  \Delta S = 0,1; & g_{w.el} = 0, \\
0.72 & \mbox{ if } &  \Delta S = 0;   & g_{w.el}\neq 0 ,
\end{array}
\label{e21}
\end{equation}
when in addition to $A_{0}$ only one of $A_{i}$, $i=1,2$ or
$A_{g}(M1 \mbox{or} E1)$
is taken into account (we put also $A_{3}=0$ in the following).Solving a
system of two equations, the first one being the definition of $F/D_{ax}$ in
terms of $A_{0}$ and $A_{i}$ or, alternatively, in terms of $A_{0}$ and
$A_{g}$, while the second one is the normalization condition, we obtain the
values collected in Table 1.
\begin{center} Table 1.      \\
\end{center}
\begin{center}
\begin{tabular}{lccccc}
& & & & &  \\
${F\over D}$ & $A_0^2$ &
$A_1^2$ & $A_2^2$ &  $A_g^2(M1)$  & $A_g^2(E1)$ \\
& & & & &  \\
0.58 & 0.81 & -- & 0.19 & -- & -- \\
& & & & &  \\
0.58 & 0.35 & -- & -- &  0.65 & --  \\
& & & & & \\
0.72 & 0.86 & 0.14 & -- & -- & -- \\
& & & & &  \\
0.72 & 0.51 & --  & -- & 0.49 & --  \\
& & & & & \\
$0.72|_{ax}^{mag}$ & 0.72 & 0.114 & 0.005 & -- & 0.16  \\
& & & & & \\
$0.687|_{QM}$ & 0.938 & 0.059 & 0.002 & -- & --  \\
& & & & & \\
 \end{tabular}\\
\end{center}
Too large values of either $A_{i}$ or $A_{g}$ on the first four lines
look rather difficult to accept. The entries on the 5th line correspond to
solution of the enlarged system of equations\cite{a_9} including $F/D|_{mag}$,
which is expressed in terms of $A_{i} {,} i=0,1,2$ and $A_{g}$.  Although
appearing to be more attractive, they deviate from typical values of
nonrelativistic quark model (QM) represented on the 6th line.The QM-results
are obtained via diagonalization of the hamiltonian containing spin-dependent
potentials induced by the one-gluon exchange \cite{16}.Thus, the increase of
the spin-tensor (spin-spin) potential is responsible for the larger value of
$A_{2}$ ($A_{1}$) and, respectively, for the decrease ( or increase) of
$F/D$-ratio compared to $2/3$.  Concerning the magnitude of $A_{g}^2$, we note
that $A_{0}^2 |_{QM}$ is the sum of two terms
\begin{equation}
0.938 = A_{0}^{2}({\underline {56}}_{S}) + A_{0}^{2}({\underline {56}}_{S^{'}}) \simeq
0.85 + 0.09,
\label{e22}
\end{equation}
Therefore, if the Roper resonance
$N^{\ast}(1440)$ and other members of lowest $J^{P}= {1\over 2}^{+}$-
multiplet, traditionally considered to be the radially excited
${\underline{56}}_{S^{'}}$- multiplet, would largely to be hybrid states, we
should exclude their contributions to $A_{0}^2$ in favour of either of
$A_{g}^2$. In that case the mentioned discrepancy is diminished.\\

{\bf 3}. We conclude with the following remarks:\\

1) The deviation of the ratio $F/D=.72$, Eq.(12), from the $SU(6)$ --value
$2/3$ shows, that despite the validity of the celebrated
$SU(6)$--ratio  $\mu(P)/\mu(N)=-3/2$, the $SU(6)$--symmetry
is strongly broken. The importance of taking into account the nonvalence
degrees of freedom in relevant parametrization of the observables
within the (broken) internal symmetries is demonstrated.\\

2) The new more accurate angular correlation measurements in different
$Y \to Nl\nu$-- decays could give, as exemplified in \cite{a_13},
very important information on second--class currents and new values of
the $g_{A}$.\\

3) The current and forthcoming measurements of the flavour-separated
$\Delta q$'s are of indispensable value to discriminate between different
approaches to the spin phenomena at high energies.\\

This work was supported by the Russian Foundation for Basic Researches,
grants No. 96-02-18137 and 96-15-96423.

\end{document}